\documentclass[12pt]{article}

\def\beq{\begin{equation}}
\def\eeq{\end{equation}}
\def\beqn{\begin{eqnarray}}
\def\eeqn{\end{eqnarray}}

\usepackage{amsmath,amssymb}
\usepackage{braket}
\usepackage[matrix, arrow, curve]{xy}
\usepackage{graphicx}
\begin{document}
\begin{titlepage}

\begin{flushright}
ITEP-TH- 02/14\\
\end{flushright}

\vspace{1cm}

\begin{center}
{  \Large \bf RG limit cycles}
\end{center}
\vspace{1mm}

\begin{center}

 {\large
 \bf  K.Bulycheva$\,^{a}$, A.Gorsky$\,^{a,b}$ }

\vspace{3mm}

$^a$
{\it Institute of Theoretical and Experimental Physics,
Moscow 117218, Russia}\\[1mm]
$^b$
{\it Moscow Institute of Physics and Technology,
Dolgoprudny 141700, Russia}
\end{center}

\centerline{\small\tt gorsky@itep.ru }

\centerline{\small\tt bulycheva@itep.ru }
\vspace{1cm}
\centerline{ Contribution to the "Pomeranchuk-100" Volume}
\vspace{2cm}
\centerline {\large\bf Abstract}
  \vspace{1cm}
In this review we consider the concept
of limit cycles in the renormalization group flows. The
examples of this phenomena in the quantum mechanics
and field theory will be presented.


\end{titlepage}

\section{Generalities}

It is usually assumed that the RG flow connects
fixed points, starting at a UV repelling point and terminating at a IR attracting point. However it turned out that this open
RG trajectory does not exhaust all possibilities and
the clear-cut quantum
mechanical example of the nontrivial  RG limit cycle has been found in \cite{glazek}
confirming the earlier expectations. This example triggered the search
for patterns of this phenomena which was quite successful.
They
have been identified both in the systems with finite number of degrees of freedom \cite{rd, braaten,bavin,beane}
and in the field theory framework \cite{leclair, son,gorsky}. Now the cyclic
RG takes its prominent place in the world of RG phenomena however
the subject certainly deserves much more study.

The appearance of critical points corresponds to phase transitions of the second kind, hence there exists a natural question concerning the connection of RG cycles with phase transitions. The very phenomenon of the cyclic RG flow has been interpreted in the
important paper \cite{son} as a kind
of generalization of the BKT phase transitions in two dimensions. One can start from a usual example of an RG flow connecting UV and IR fixed points and then consider a motion in a parameter space which results in a merging of the fixed points.In \cite{son} it was argued that when the parameter goes into the complex region the cyclic behaviour of the RG flow gets manifested and a gap in the spectrum arises.
This happens similar to the BKT
transition  case when a deconfinement of vortices occurs at the critical temperature
and the conformal symmetry is restored at  lower temperatures.
The appearance of the RG cycles can be also interpreted as the peculiar
anomaly in the  classical conformal group \cite{anomaly}. This anomaly
has the origin in some "falling to the center" UV phenomena which
could have quite different reincarnations.
We would like to emphasize one more generic feature of the phenomena ---
the cyclic RG usually occurs in the system with at least two couplings. One of them
undergoes the RG cyclic flow while the second determines the period of the cycle.

The collision of the UV and IR fixed points can be illustrated in a quite general
manner as follows. Assume that there are two couplings
$(\alpha, g)$ in the theory and we focus at the renormalization
of the coupling which enjoys  the following $\beta$-function

\beq
\beta_g = (\alpha -\alpha_0) - (g-g_0)^2,
\label{beta}
\eeq
which vanishes at the hypersurface in the parameter space

\beq
g=g_0 \pm \sqrt{\alpha-\alpha_0}.
\eeq

It was argued  in \cite{son} that the collision of two roots at $\alpha =\alpha_0$
can be interpreted as the collision of UV and IR fixed points.
Upon the collision the points move into the complex $g$ plane and an
RG cycle emerges. The period of the cycle
can be immediately estimated as

\beq
T\propto \int_{g_{UV}}^{g_{IR}} \frac{dg}{\beta(\alpha;g)} \propto \frac{1}{\sqrt{\alpha-\alpha_0}}.
\eeq
The phenomena is believed to be generic once the beta--function has the form (\ref{beta}).
Note that is was shown that the RG cycles are consistent with the c-theorem \cite{zachos}.

Breaking of the conformal symmetry results in
the generation of the mass scale which has non-perturbative nature.
Due to the RG cycles the scale is not unique and the whole tower with the
Efimov-like scaling gets manifested
\beq
E_{n+1}=\lambda E_n,
\eeq
where $\lambda$ is fixed by the period of RG cycle.

In the examples available we could attempt to trace the physical picture
behind. It turns out that the origin of two couplings is quite general. One coupling
does not break the conformal symmetry which is exact in some subspace of the parameter
space. The second coupling plays the role of UV regularization which can be imposed
in one or another manner. It breaks the conformal symmetry however some discrete
version of the scale symmetry survives which is manifested in the cycle structure.
The UV regularization will have different reincarnations in the examples considered:
the account of the finite size of the nuclei, contact interaction in the
model of superconductivity or the brane splitting in the supersymmetric models.

Historically the first example of this phenomena has been found
long time ago by Efimov \cite{efimov} in the context of nuclear physics.
He considered the three-body system when two particles are near
threshold and have attractive potential with the third particle.
It was shown  that two--particle bound states are absent in the spectrum, but there is a tower of the three--particle bound states with
the geometrical  scaling corresponding to $\lambda \approx 22,7$.
The review of the RG interpretation of the Efimov
phenomena can be found in \cite{hammer}.

When considering the system with finite number of degrees of freedom
the meaning of the RG flows has to be clarified. To this aim
some UV cutoff should be introduced. In the first
example in \cite{glazek} the step of the RG corresponds to the
integrating out the highest energy level taking into account its
correlation with the rest of the spectrum.  This
approach has a lot in common with the renormalization procedure in the
matrix models considered in \cite{brezin}. The same UV cutoff for
formulation of RG procedure
has been used for the Russian Doll (RD) model describing the restricted
BCS model of superconductivity \cite{lrs}. In that case the coupling providing
the Cooper pairing undergoes the RG cycle while the CP-violating
parameter defines the period.

In the second class of examples the UV cutoff is introduced
not at high energy scale but at small distances.
The RG cycles have been
found in the non-relativistic Calogero-like models with $\frac{1}{r^2}$
potential  which enjoys naive conformal symmetry \cite{braaten,bavin,beane}.
The RG flow is formulated in terms of the short distance regularization of the model.
It is assumed that the wave function with $E=0$ at large $r$ does not depend
on the UV cutoff at small $r$. This condition yields the equation for the parameter
of a cutoff in the regularization potential. This equation has multiple solutions which can be
interpreted as the manifestation of the tower  of shallow bound states  with
the Efimov scaling in the regularized Calogero model with attraction.
The scaling factor in the tower  is determined by the Calogero coupling constant
which reflects the remnant of the conformal group upon the regularization.

The list of the field theory examples in  different dimensions with the
cyclic RG flows is short but quite representative. In two dimensions
the explicit example with the RG cycle has been found in some range
of parameters in the sin-Gordon model.
The cycle manifests itself
in the pole structure of the $S$-matrix. Efimov-like tower of states
corresponds to the specific poles with the Regge-like
behavior of the resonance masses  \cite{leclair}
\beq
m_n=m_s e^{\frac{n\pi}{h}},
\eeq
where $h$ is a certain parameter of the model.
Moreover it was argued that the
$S$-matrix behaves universally under the cyclic RG flows. The tower of Efimov states scales in the same manner as in the quantum mechanical case.

The origination of the cyclic RG behavior in the sine--Gordon model is not
surprising.
Indeed it was argued in \cite{son} that the famous Berezinsky--Kosterlitz--Thouless (BKT) phase transition
in XY system belongs to this universality class. On the other hand  one can map the XY system
at the $T$ temperature into the
sin-Gordon theory with the parameters:
\beq
L_{SG}= T(\partial \phi) ^2 - 4z \cos\phi,
\eeq
and look at the renormalization of the interaction coupling. The $\beta$--functions
read as
\beq
\beta_u = -2v^2, \qquad \beta_v= -2uv,
\eeq
where
\beq
u= 1-  \frac{1}{8\pi T}, \qquad v= \frac{2z}{T\Lambda^2},
\eeq
and $\Lambda$ is the UV cutoff introduced to regularize the vortex core.
The form of $\beta$ functions implies the existence of the limit cycle with the following
expression for the correlation length:
\beq
\xi_{BKT}\Lambda \propto \exp\left(\frac{c}{\sqrt{|T-T_c|}}\right),
\eeq
above the phase transition. This RG behavior gets mapped
into the RG cycle in the sine--Gordon model.

The example of the Efimov tower in 2+1 dimensions has been found
in \cite{sonvletnyuyunoch} in the holographic representation. The
model is based on the $D3-D5$ brane configuration and corresponds
to the large $N$ 3d gauge theory with fundamentals enjoying  $\mathcal{N}= 4$
supersymmetry. In addition the magnetic field and the finite density
of conserved charge are present. At  strong coupling the gauge theory
is described in terms of the probe $N_f$ flavor branes in the
nontrivial $AdS_5 \times S^5$ geometry when the $U(1)$ bulk gauge field
is added providing the magnetic field in the boundary theory.

The generation of the tower of the Efimov states happens at some
value of the ``filling fraction'' $\nu$ in external magnetic field.
The phase transition corresponds to the change of the minimal
embedding of the probe $D5$ branes in the bulk geometry with
the BKT critical behavior of the order parameter. In that case
the order parameter gets identified with the condensate $\sigma$
which behaves as:
\beq
\sigma \propto \exp\left(-\frac{1}{\nu}\right).
\eeq
Above the phase transition the embedding gets changed and the brane
becomes extended in one more coordinate. The scale associated
with this extension into new dimension is nothing but the nonperturbative
scale amounting to the mass gap.
The phenomena of the cyclic RG flow in this case
has the Breitenlohner-Freedman instability as the gravitational counterpart.

In four dimensions the most famous example of the Efimov tower is
the so-called Miransky scaling for the condensate in the magnetic
field. In \cite{miransky} was argued that the chiral condensate is generated
in the external magnetic field in the abelian theory with the following behavior:
\beq
\langle\bar{\Psi} \Psi\rangle\propto \Lambda^3 \exp(-\frac{c}{\sqrt{\alpha - \alpha_{crit}}}),
\eeq
where $\alpha$ is the fine--structure constant, and $c$ is some parameter of the model.

More recent example \cite{kiritsis} of the Efimov tower in four dimensions concerns
the Veneziano limit of QCD when $N_f,N_c \rightarrow \infty$ while
the ratio $x= \frac{N_f}{N_c}$ is fixed. It turns out that this parameter
can be considered as the variable in the RG flow which reminds the finite-dimensional
examples. At some value of  RG scale the tower of condensates gets generated
with geometrical Efimov scaling. The period of the RG cycle reads as:
\beq
T\propto \frac{\kappa}{\sqrt{x_c-x}},
\eeq
where $x_c$ is the critical value of the $x$ parameter.
Finally the 4d example with the RG cycle has been found in the $\mathcal{N}=2$
SUSY gauge theory  in the $\Omega$-background \cite{gorsky}. In that case
the gauge coupling undergoes the RG cycle whose period is determined
by the parameter of the $\Omega$-background,
\beq
T\propto \epsilon^{-1}.
\eeq
The appearance of the RG cycle in this model can be traced from its
relation with the quantum integrable systems of the spin chain type.

In this review we provide the reader with the
examples of this phenomenon. The list of the systems with finite
number of degrees of freedom involves the Calogero model and the relativistic
model with the classical conformal symmetry describing the external
charge in graphene. Another finite-dimensional example concerns the RD
model of the restricted BCS superconductivity. The field theory
examples concern the $3d$ and $4d$ theories in external fields.
We shall focus on their brane representations and use 
their relations to the finite dimensional integrable systems.

\section{RG cycles in non-relativistic quantum mechanics}
\label{RG_Cal}

In this Section we consider the example of the limit cycle
in RG in the non-relativistic system with the inverse-square potential, or the Calogero system:

\begin{equation}
    H=\frac{\partial^2}{\partial r^2} -\frac{\mu(\mu-1)}{r^2}.
    \label{ham_cal}
\end{equation}

The distinctive feature of the system described by the Hamiltonian (\ref{ham_cal}) is its conformality. Namely, the operators $(H,D,K)$, where $D$ is the dilation generator and $K$ is the conformal boost, generate the conformal $\mathfrak{sl}_2$ algebra (see  Section \ref{sec:anomaly}).

The eigenfunctions of (\ref{ham_cal}) having finite energy immediately break this symmetry; more non-trivial is the fact that even the ground state breaks conformality. Namely, the solution to the $H\psi=0$ equation is the following:

\begin{equation}
    \psi_0=c_+r^\mu+c_-r^{1-\mu}.
    \label{sol_cal}
\end{equation}
This solution is scale-invariant only if one of the coefficients $c_\pm$ is zero. If both the coefficients are present, they define an intrinsic length scale $L=\left( c_+/c_- \right)^{1/(-2\mu+1)}$. Requiring that the quantity $c_+/c_-$ which describes the ground-state solution be invariant under the change of scale,

\begin{equation}
    \frac{c_+}{c_-}=-r_0^{-2\mu+1}\frac{\gamma-\mu+1}{\gamma+\mu},
    \label{c_rg}
\end{equation}
we arrive at the beta-function for the $\gamma$ parameter,

\begin{equation}
    \beta_\gamma=\frac{\partial\gamma}{\partial\log r_0}=-\left( \gamma+\mu \right)\left( \gamma-\mu+1 \right)=\left( \mu-\frac12 \right)^2-\left( \gamma-\frac12 \right)^2,
    \label{gamma_rg}
\end{equation}
where $r_0$ is the RG scale. We can identify $\gamma=\mu-1, \gamma=-\mu$ points, i.e. solutions with $c_+=0$, $c_-=0$, with UV and IR attractive points of the renormalization group flow \cite{son}.

If $\mu=i\nu$ is imaginary, i.e. the potential is attractive, then the equation (\ref{gamma_rg}) allows us to determine the period of the renormalization group:

\begin{equation}
    T=-\int_{-\nu+1}^{\nu}\frac{d\gamma}{\beta_\gamma}=\frac{\pi}{\nu-\frac12}.
    \label{T_rg}
\end{equation}

This means that an infinite number of scales is generated, differing by a factor of $\exp\left( -\frac{\pi}{\nu-\frac12} \right)$. To see this explicitly, we find the solutions to the Schr\"odinger equation at finite energies. In the attractive potential the solution (\ref{sol_cal}) can be written as:

\begin{equation}
    \psi_0\propto \sqrt{r}\sin\left( \left( \nu-\frac12 \right)\log \left( \frac{r}{r_0} \right) +\alpha \right).
    \label{sol_cal_im}
\end{equation}
We observe that this solution oscillates indeterminately in the vicinity of the origin and there is no way to fix the $\alpha$ constant. To regularize this behaviour, we can break the scale invariance at the level of the Hamiltonian and introduce a regularizing potential. Two most popular regularizations involve
the square-well potential \cite{bavin,beane} or the delta-shell potential \cite{braaten}. One more choice is to introduce a $\delta$-function at the origin \cite{son}.

Choosing the square-well regularization,

\begin{equation}
    V(r)=\left\{
    \begin{array}{l}
        -\frac{\nu(\nu-1)}{r^2}, r>R,\\
        -\frac{\lambda}{R^2}, r\le R,
    \end{array}
    \right.
    \label{V_cal}
\end{equation}
we require that the action of the dilatation operator on the wavefunction inside the well and outside it be equal at $r=R$. This condition amounts to the equation on $\lambda$,

\begin{equation}
    \sqrt{\lambda}\cot\sqrt{\lambda}=\frac12+\nu\cot\left( \nu\log\left( \frac{R}{r_0} \right) \right).
    \label{lambda_rg}
\end{equation}
The multivalued function $\lambda(R)$ can be chosen to be continuous \cite{beane}.

The wavefunction regular at infinity is given as a combination of the Bessel functions \cite{beane},

\begin{equation}
    \psi\left( r,\kappa_m \right)=\sqrt{r}(-1)^m\left( ie^{-i\nu\frac{\pi}{2}}J_{i\nu}\left( \kappa_m r \right)-ie^{i\nu\frac{\pi}{2}}J_{-i\nu}\left( \kappa_m r \right), \right),
    \label{psi_cal}
\end{equation}
where $\kappa$ is the energy of the state. The spectrum consists of infinitely many shallow bound states with adjacent energies differing by an exponential factor,

\begin{equation}
    \frac{\kappa_{m+1}}{\kappa_m}=e^{-\frac{\pi}{\nu}}.
    \label{kappa1}
\end{equation}

Note that the coordinate enters the wavefunction (\ref{psi_cal}) only in combination with energy, and the spectrum is generated by the dilation operator:

\begin{equation}
    \psi_{m+1}=\exp\left( -\frac{\pi}{\nu}r\partial_r \right)\psi_m.
    \label{spectrum_gen}
\end{equation}

One can think of that relation as that the action of the dilatation operator shifts zeroes of the wave function from the area of $r<R$ to the area with the inverse square potential, and one step of (\ref{spectrum_gen}) evolution corresponds to elimination of a single zero in the area with the square-well potential. Since the wave function oscillates infinitely at the origin, the elimination of all the zeroes would require an infinite number of steps, and in this way a whole tower of states gets generated.

\section{RG cycle  in graphene}
In this Section we shall consider the similar problem in 2+1 dimensions
which physically corresponds to the external charge in the planar graphene layer.
The problem has the classical conformal symmetry and is the relativistic  analogue
of the conformal non-relativistic Calogero-like system. Due to  conformal symmetry
we could expect the RG cycles and Efimov-like states in this problem
upon imposing the short distance cutoff.
The issue of the charge in the graphene plane has been discussed
theoretically \cite{levitov2,gra1,gra2} and experimentally \cite{exp1,exp2}. It was
argued that indeed there is the tower of "quasi-Rydberg" states with the exponential
scaling \cite{levitov}. The situation can be interpreted as an atomic collapse phenomena similar
to the instability of $Z>137$ superheavy atoms in QED \cite{atoms}.

Turn now to the consideration of an electron in graphene which interacts with
an external charge. The two-dimensional Hamiltonian reads as,
\beq
H_D = v_{F} \sigma_i p^i +V(r), \qquad i=1,2.
\label{dirac_ham}
\eeq
The external charge creates a Coulomb potential,
\beq
V(r) = -\frac{\alpha}{r}, \qquad r\ge R.
\label{pot_coul}
\eeq
As we shall see, the solution in presence of the potential (\ref{pot_coul}) oscillates indefinitely at the origin and needs to be regularized by some cutoff $R$. Hence close enough to the origin $r\le R$ the potential (\ref{pot_coul}) gets replaced by some constant potential $V_{reg}(r,\lambda(R))$.
The renormalization condition for the $\lambda$ parameter is that the zero-energy wave function is not dependent on the short-distance regularization. This condition is chosen similarly to that of the renormalization of the Calogero system (see Section \ref{RG_Cal}). Hence our primary task is to find the zero-energy solution to the Dirac equation,

\begin{equation}
    H_D\psi_0=0.
    \label{dirac}
\end{equation}

Since the Hamiltonian commutes with the $J_3$ operator,

\begin{equation}
    J_3=i\frac{\partial}{\partial \varphi}+\sigma_3, \qquad \left[ H_D, J_3 \right]=0,
    \label{J3}
\end{equation}
we can look for the solutions of (\ref{dirac}) in the form:

\begin{equation}
    \psi_0=\begin{pmatrix}\chi_0(r)\\ \xi_0(r)e^{i\varphi}  \end{pmatrix}, \qquad J_3\psi_0=\psi_0.
    \label{spinor}
\end{equation}

Then in polar coordinates the equation (\ref{dirac}) reads as:

\begin{equation}
    \left\{
        \begin{array}{l}
            -i\hbar v_F \left( \partial_r+\frac 1r \right)\xi_0  = -V(r) \chi_0,\\
            -i\hbar v_F \partial_r \chi_0= -V(r)\xi_0,
        \end{array}
    \right.
    \label{dirac1}
\end{equation}
which is equivalent to:

\begin{equation}
    \left\{
    \begin{array}{l}
        \xi_0(r)=i\hbar v_F (V(r))^{-1}{\partial_r \chi_0},\\
        \partial_r^2 \chi_0+ \left( \frac{1}{r}-\frac{V'(r)}{V(r)} \right)\partial_r\chi_0 +\frac{V^2(r)}{\hbar^2 v_F^2} \chi_0 = 0.
    \end{array}
    \right.
    \label{Dechi}
\end{equation}

For the potential $V =- \frac{\alpha}{r}$ we get the following equation on $\chi_0(r)$:

\begin{equation}
    \partial_r^2\chi_0+\frac{2}{r}\partial_r\chi_0+\frac{\beta^2}{r^2}\chi_0=0, \qquad \beta=\frac{\alpha}{\hbar v_F}.
    \label{eq_coul}
\end{equation}
Supposing that $\beta^2 = \frac{1}{4} + \nu^2$ we write the solution as:

\begin{equation}
    \chi_0= \sqrt{r} \left(c_- \left( \frac{r}{r_0} \right)^{-i\nu}+c_+ \left( \frac{r}{r_0} \right)^{i\nu}  \right)\propto \sqrt{r}\sin\left( \nu \log \frac{r}{r_0}+\varphi \right).
    \label{sol_coul}
\end{equation}

We see that this solution shares the properties of the ground-state Calogero wavefunction (\ref{sol_cal}), namely at nonzero $c_{\pm}$ it generates its own intrinsic length scale and it oscillates indeterminately at the origin. In order to fix the $\varphi$ constant we need to introduce a cut-off potential. Hence we consider the solution in the potential:

\begin{equation}
    V(r)=\left\{
    \begin{array}{l}
        -\frac{\alpha}{r}, \qquad r>R,\\
        V_{reg}=-\hbar v_F \frac{\lambda}{R}, \qquad r\le R.
    \end{array}
    \right.
    \label{reg_pot}
\end{equation}
The dilatation operator acts on $\chi$ as following:

\begin{equation}
    r\partial_r \chi_0=\left( \frac 12 +\nu \cot \left( \nu\log \frac{r}{r_0} \right) \right)\chi_0.
    \label{dil_coul}
\end{equation}
For the constant potential $V_{reg}$ we get from $(\ref{Dechi})$:

\begin{equation}
    \partial^2_r \chi_0^{reg}+\frac{1}{r}\partial_r \chi_0^{reg} + \frac{\lambda^2}{R^2} \chi_0^{reg} = 0.
    \label{eq_const}
\end{equation}
Choosing the solution of (\ref{eq_const}) which is regular at the origin we obtain,

\begin{equation}
    \chi_0^{reg}\propto J_0\left(\lambda \frac{r}{R} \right).
    \label{sol_const}
\end{equation}
Computing the action of the dilation operator on the solution in the area of constant potential and equating it to the action of the dilation operator (\ref{dil_coul}) we get the equation on the $\lambda$ regulator parameter:


\begin{equation}
    \frac 12+ \nu\cot\left( \nu\log\left( \frac{R}{r_0} \right) \right)=-\lambda\frac{J_1(\lambda)}{J_0(\lambda)}.
    \label{rg_graph}
\end{equation}

The equation (\ref{rg_graph}) defines $\lambda$ as a multi-valued function of $R$. The period of the RG flow corresponds to jump from one branch of the $\lambda(R)$ function to another.

Now we proceed to find the bound states in the (\ref{pot_coul}) potential. We consider again the Dirac equation,

\begin{equation}
    H_D\psi_\kappa=-\hbar v_F \kappa \psi_\kappa.
    \label{dirac_E}
\end{equation}
Then the equation on $\chi$ analogous to (\ref{Dechi}) is as following:

\begin{equation}
    \partial^2_r\chi_\kappa+\frac{2\beta-\kappa r}{\beta-\kappa r}\frac{1}{r}\partial_r\chi_\kappa+\left( \frac{\beta}{r}-\kappa \right)^2\chi_\kappa=0.
    \label{eq_coul_E}
\end{equation}
Asymptotically when $r\gg \frac{\beta}{\kappa}$ the solution of (\ref{eq_coul_E}) regular at infinity is given by the Hankel function,

\begin{equation}
    \chi_\kappa\propto H_0^{(1)}(i\kappa r).
    \label{sol_inf}
\end{equation}

At small $r\ll \frac{\beta}{\kappa}$ the solution is not regular at the origin,

\begin{equation}
    \chi_\kappa\propto \sqrt{r}\sin\left( \nu \log \frac{r}{r_0} \right),
    \label{sol_E}
\end{equation}
and we are again in need for the regulator potential. Solving again the Dirac equation (\ref{dirac_E}) in presence of the constant potential $V_{reg}$ and computing the action of the dilatation operator,

\begin{equation}
    r\partial_r \chi_k^{reg}=-\left( \lambda-\kappa R \right)\frac{J_1\left(\lambda-\kappa R \right)}{J_0\left( \lambda-\kappa R \right)} \chi_\kappa^{reg},
    \label{dil_reg_E}
\end{equation}
we can equate (\ref{dil_reg_E}) to the action of the dilatation operator on (\ref{sol_E}) and get the equation on the spectrum of the bound states,

\begin{equation}
    \frac 12+\nu\cot\left( \nu\log\left( \kappa R \right) \right)=-\left( \lambda-\kappa R \right)\frac{J_1\left( \lambda-\kappa R \right)}{J_0\left( \lambda-\kappa R \right)}.
    \label{rg_kappa}
\end{equation}

This condition gives the spectrum of infinitely many shallow bound states,

\begin{equation}
    \kappa_n=\kappa_* \exp\left( -\frac{\pi n}{\nu} \right), \qquad \kappa\to \infty.
    \label{kappa}
\end{equation}

\section{Anomaly in the ${\bf so}(2,1)$ algebra}
\label{sec:anomaly}
Let us make some comments on the algebraic counterpart
of the phenomena considered following \cite{anomaly}.
As we have mentioned the
conformal symmetry is the main player since Hamiltonians under
consideration are  scale invariant before regularization.
Actually this group can be thought of as the example of
spectrum generating algebra when the Hamiltonian is one of
the generators or is expressed in terms of the generators
in a simple manner. This is familiar from the exactly or
quasi-exactly solvable problems when the dimension
of the representation selects the size of the algebraic part of the
spectrum.

Let us introduce the generators of the ${\bf so}(2,1)$ conformal
algebra $J_1, J_2, J_3$:
the Calogero Hamiltonian,
\beq
J_1= H= p^2 +V(r),
\eeq
the dilatation generator,
\beq
J_2=D = t H -\frac{1}{4}(pr + r p),
\eeq
and the generator of special conformal transformation,
\beq
J_3=K =  t^2 H - \frac{t}{2}(pr +r p) +\frac{1}{2}r^2.
\eeq
They satisfy the relations of the ${\bf so}(2,1)$ algebra:
\beq
[J_2,J_1]= -iJ_1,\qquad [J_3,J_1]= -2iJ_2,\qquad [J_2,J_3]= iJ_3.
\eeq

The singular behavior of the potential at the origin
amounts to a kind of anomaly in  the ${\bf so}(2,1)$ algebra,
\beq
A(r)=-i[D,H]+H,
\eeq
which in $d$ space dimensions can be presented in the
following form:
\beq
A(r)= -\frac{d-2}{2}V(r) + (r^i\nabla_i) V(r).
\label{simple}
\eeq
The simple arguments imply the following relation
\beq
\frac{d}{dt}\langle D\rangle_{\mathrm{ground}}=E_{\mathrm{ground}},
\label{anomaly}
\eeq
where the matrix element is taken over the
ground state.

It turns out that (\ref{anomaly}) is fulfilled
for the singular potentials in Calogero-like model
or in models with contact potential, $V(r)=g\delta(r)$. The expression for anomaly
does not depend on the regularization chosen.
Moreover more detailed analysis demonstrates that
the anomaly is proportional to the $\beta$-function
of the coupling providing the UV regularization
as can be expected.

A similar calculation of the anomaly for the graphene
case can be performed for arbitrary state,
\beq
\left\langle{\frac{d D}{d t}}\right\rangle_\psi = \braket{\Xi}_\psi =  - \int d^2 x \psi^*(V(x)+x_i \partial_i V(x)) \psi,
\eeq
which yields using square-well regularization:
\beq
\braket{\Xi}_\psi =\hbar v_F \frac{\lambda(R)}{R} \frac{\int \limits^{R}_0 r |\psi|^2 dr}{\int \limits^{\infty}_0 r |\psi|^2 dr}.
\eeq

It is convenient to use the two-dimensional identity in (\ref{simple}),
\beq
\nabla \frac{\vec{r}}{r}= 2\pi \delta(\vec{r}),
\eeq
which simplifies the calculation of the anomaly for any normalized
bound state,
\beq
\frac{d}{dt}\langle D\rangle_{\Psi} = -g \pi \int d^2 r \delta (r)|\Psi(r)|^2.
\eeq

\section{RG cycles in models of superconductivity}
\label{rg_chain}
In this Section we explain how the cyclic RG flows emerge in
truncated models of superconductivity. To this aim we shall first
describe the Richardson model and then consider its generalization
to the RD model which enjoys the cyclic RG flow. These models are distinguished by the finiteness of the number of fermionic levels. The relation with
the integrable many-body systems proves to be quite useful.

\subsection{Richardson model versus Gaudin model}

Let us recall the truncated BCS-like Richardson model of superconductivity \cite{richardson}
with some number
of doubly degenerated fermionic levels with the energies $\epsilon_{j\sigma}, j=1,\dots, N$.
It describes the system
of a  fixed number of the Cooper pairs.
It is assumed that  several energy levels are populated by  Cooper pairs while
levels with the single fermions are blocked.
The Hamiltonian reads
as
\beq
H_{BCS}= \sum_{j,\sigma= \pm}^N \epsilon_{j\sigma} c^{+}_{j\sigma}c_{j\sigma} -
G\sum_{jk}c^{\dagger}_{j+}c^{\dagger}_{j -}c_{k -}c_{k+},
\eeq
where $c_{j\sigma}$ are the fermion operators and $G$ is the
coupling constant providing the attraction leading to the formation
of the Cooper pairs. In terms of the hard-core boson operators
it reads as
\beq
H_{BCS}=\sum_j\epsilon_j b^{\dagger}_jb_j - G \sum_{jk}b^{\dagger}_jb_k,
\eeq
where
\beq
[b^{\dagger}_j,b_k]= \delta_{jk}(2N_j-1), \qquad b_j=c_{j -}c_{j +},\qquad N_j= b^\dagger_jb_j.
\eeq

The eigenfunctions of the Hamiltonian can be written as,
\beq
|M\rangle=\prod_i^M B_i(E_i)|\mathrm{vac}\rangle, \qquad B_i= \sum _j^N \frac{1}{\epsilon_j- E_i} b^{\dagger}_j,
\eeq
provided the Bethe ansatz equations are fulfilled,
\beq
\label{BA}
G^{-1}= - \sum _j^N \frac{1}{\epsilon_j- E_i} + \sum _j^M \frac{2}{E_j- E_i}.
\eeq
The energy of the corresponding states reads as:
\beq
E(M)= \sum_i E_i.
\eeq

It was shown in \cite{richsolution} that the Richardson
model is exactly solvable and closely related to the particular generalization
of the Gaudin model. To describe this relation it is convenient
to introduce the so-called pseudospin ${\bf sl}(2)$ algebra in
terms of the creation-annihilation
operators for the Cooper pairs,
\beq
t_j^{-}=b_j, \qquad t_j^{+}=b^\dagger_j, \qquad t^{0}_j=N_j-1/2.
\eeq

The Richardson Hamiltonian commutes with
the set of operators $R_i$,
\beq
R_i= -t^0_i -2G\sum^N_{j\neq i}\frac{t_i t_j}{\epsilon_i -\epsilon_j},
\eeq
which are identified as the Gaudin Hamiltonians,
\beq
[H_{BCS},R_j]=[R_i,R_j]=0.
\eeq
Moreover the Richardson Hamiltonian itself
can be expressed in terms of the operators $R_i$ as:
\beq
H_{BCS}= \sum_i \epsilon_i R_i + G\left( \sum R_i \right)^2 + \mathrm{const}.
\eeq

The number $N$ of the fermionic levels coincides
with the number
of sites in the Gaudin model and the coupling constant in the
Richardson Hamiltonian corresponds to the "twisted boundary
condition" in the Gaudin model. The Bethe ansatz equations
for the Richardson model (\ref{BA}) exactly coincide with
the ones for the generalized Gaudin model. It was argued in \cite{rd} that the
Bethe roots correspond to the excited Cooper pairs that
is natural to think about the solution to the Baxter
equation as the wave function of the condensate. In terms
of the conformal field theory Cooper pairs correspond
to the screening operators \cite{sierraconf}.

For the nontrivial degeneracies of the energy levels $d_j$
the BA equations read as:
\beq
G^{-1}= - \sum _j^N \frac{d_j}{\epsilon_j- E_i} + \sum _{j\neq i}^M \frac{2}{E_j- E_i}.
\eeq

\subsection{Russian Doll model of superconductivity and  twisted XXX spin chains}

The important generalization of the Richardson model
describing superconductivity
is the so-called RD model \cite{rd}. It involves the
additional dimensionless parameter $\alpha$ and the RD Hamiltonian reads as:

\begin{equation}
 H_{RD}= 2\sum_i^N (\epsilon_i- G)N_i -\bar{G}\sum_{j<k}
(e^{i\alpha} b^{+}_k b_j +e^{-i\alpha} b^{+}_jb_k),
  \label{ham_RD}
\end{equation}
with two dimensionful parameters $G, \eta$
and $\bar{G} =\sqrt{G^2 +\eta^2}$. In terms of these variables the
dimensionless parameter
$\alpha$  has the following form:
\beq
\alpha=\arctan \left( \frac{\eta}{G} \right).
\eeq
It is  also useful to consider   two dimensionless parameters $g,\theta$
defined as $G=gd$ and $\eta =\theta d$ where $d$ is the level spacing.
The RD model reduces to the Richardson model in the limit $\eta\rightarrow 0$.

The RD model turns out to be integrable as well. Now instead
of the Gaudin model the proper counterpart
is  the generic quantum twisted XXX spin chain \cite{rdsolution}.
The transfer matrix of such spin chain
model $t(u)$ commutes with the $H_{RD}$ which itself can be
expressed in terms of the spin chain Hamiltonians.

The equation defining the spectrum of the RD model reads as:

\begin{equation}
        e^{2i\alpha}\prod_{l=1}^N\frac{E_i-\varepsilon_l+i\eta}{E_i-\varepsilon_l-i\eta}=\prod_{j\neq i}^M\frac{E_i-E_j+2i\eta}{E_i-E_j-2i\eta},
            \label{bae_RD}
\end{equation}
and coincides with the BA equations for the spin chain.

Taking the logarithm of the both sides of the equation (\ref{bae_RD}) we obtain:

\begin{equation}
    \alpha+\pi Q_i +\sum_{l=1}^N \arctan \left( \frac{\eta}{E_i-\varepsilon_l} \right)-\sum_{j=1}^M \arctan\left( \frac{2\eta}{E_i-E_j} \right)=0.
    \label{bae_arctan}
\end{equation}
Note that here we have added an arbitrary integer term to account for generically multivalued arctangent function.

The RG step amounts to integrating out the $N$-th degree of freedom in the RD model, or equivalently to integrating out the $N$-th inhomogeneity in the XXX chain. This results into renormalization of the twist. From (\ref{bae_arctan}) it is easy to see that:

\begin{equation}
    \arctan \left( \frac{\eta}{G_N} \right)-\arctan \left(\frac{\eta}{G_{N-1}}\right)=\sum_{i=1}^M\arctan\left( \frac{2\eta}{E_i-\varepsilon_N} \right).
    \label{rg_arctan}
\end{equation}

When $M=1$ it implies that:

\begin{equation}
    G_{N-1}-G_N=\frac{G_N^2+\eta^2}{\varepsilon_N-G_N-E},
    \label{rg_discr}
\end{equation}
which is a discrete version of the (\ref{beta}) equation. Of course the same relation can be derived from the RD Hamiltonian (\ref{ham_RD}). If we consider the wavefunction $\psi=\sum_i^N \psi_i b_i^\dagger |0\rangle$, the Schr\"odinger equation for a state with one Cooper pair amounts to:

\begin{equation}
    \left( \varepsilon_i-G-E \right)\psi_i=(G+i\eta)\sum_{j=1}^{i-1}\psi_j+(G-i\eta)\sum_{j=i+1}^N \psi_j.
    \label{rg_schr}
\end{equation}

Integration out the $N$-th degree of freedom amounts to expressing $\psi_M$ in terms of the other modes,

\begin{equation}
    \psi_N=\frac{G+i\eta}{\varepsilon_N-G-E}\sum_{j=1}^{N-1}\psi_j,
    \label{psi}
\end{equation}
and substituting it back into the Schr\"odinger equation (\ref{rg_schr}). The $G_{N-1}$ constant in the resulting equation will differ from the initial $G_N$ value as in (\ref{rg_discr}).

 The key feature of the RD model is the multiple
 solutions to the gap equation. The
 gaps are parameterized as follows:
 \beq
 \Delta_n= \frac{\omega}{\sinh t_n}, \qquad t_n= t_0+ \frac{\pi n }{\theta}, \qquad n=0,1, \dots,
 \eeq
 where $t_0$ is the solution to the following equation:
  \beq
  \tan(\theta t_0)= \frac{\theta}{g},\qquad 0<t_0< \frac{\pi}{\theta}.
  \eeq
and $\omega =dN$ for  equal level spacing. Here $E^2=\varepsilon^2+|\Delta|^2$. This
behavior can be derived via the mean field approximation \cite{lrs}.
The gap with minimal energy defines the ground state, and the other values of the gap describe excitations. In the limit $\theta \rightarrow 0$
the gaps $\Delta_{n>0} \rightarrow 0$ and
\beq
t_0=\frac{1}{g},\qquad \Delta_0= 2\omega \exp\left(-\frac{1}{g}\right),
\eeq
therefore the standard BCS
expression for the gap is recovered. At the weak coupling
limit the gaps behave as:
\beq
\Delta_n \propto \Delta_0 e^{-\frac{n\pi}{\theta}}.
\eeq
In terms of the solutions to the BA equations the
multiple gaps correspond to the choices of the
different branches of the logarithms, i.e. to different choices of the integer $Q$ parameter in (\ref{bae_RD}).

If the degeneracy of the levels is $d_n$ then the
RD model gets modified a little bit and is related to the
higher spin XXX spin chain. The local spins $s_i$ are determined
by the corresponding higher pair degeneracy $d_i$ of the $i$-th level,
\beq
s_i=d_i/2,
\eeq
and the corresponding BA equations read as:
\begin{equation}
        e^{2i\alpha}\prod_{l=1}^N\frac{E_i-\varepsilon_l+id_l+i\eta}{E_i-\varepsilon_l-id_l-i\eta}=\prod_{j\neq i}^M\frac{E_i-E_j+2i\eta}{E_i-E_j-2i\eta}.
       \label{bae_RD_spin}
\end{equation}

\subsection{Cyclic RG flows in the RD model}

 The RD model of truncated superconductivity enjoys the cyclic RG behavior \cite{rd}.
 The RG flows can be treated as the integrating out the
 highest fermionic level with appropriate  scaling
 of the parameters using the procedure developed in
 \cite{glazek}. The RG equations read as (\ref{rg_discr}):
 \beq
 g_{N-1}=g_N + \frac{1}{N}(g_N^2 + \theta^2),\qquad \theta_{N-1}=\theta_N.
 \eeq
 At  large $N$ limit the natural RG variable is
 identified with $s=\log (N/N_0)$ and the solution to the
 RG equation is:
 \beq
 g(s)=\theta\tan\left( \theta s+\tan^{-1}\left( \frac{g_0}{\theta} \right) \right).
 \eeq
 Hence the running coupling is cyclic,
 \beq
 g(s+ \lambda) =g(s),\qquad g(e^{-\lambda}N)=g(N),
 \eeq
 with the RG period,
 \beq
 \lambda=\frac{\pi}{\theta},
 \eeq
and the total number of the independent gaps in the model is:
\beq
N_{cond}\propto \frac{\theta}{\pi} \log N.
\eeq
The multiple gaps  are the manifestations
of the Efimov-like states. The sizes of the Cooper
pairs in the $N$-th  condensates also have the
RD scaling. The cyclic RG can be derived even for the single
Cooper pair.

What is going on with the spectrum of the model during the period? It was
shown in \cite{lrs} that it gets reorganized. The RG flow
experiences discontinuities from $g=+\infty$ to $g=-\infty$
when a new cycle gets started. At each jump the lowest
condensate disappears from the spectrum,
\beq
\Delta_{N+1}(g=+\infty)= \Delta_{N}(g=-\infty),
\eeq
indicating that the $(N+1)$-th  state wave function
plays the role of $N$-th state wave function at the next cycle (see (\ref{psi})).

The same behavior can be derived from the BA equation \cite{lrs}.
To identify the multiple gaps it is necessary to remind
that the solutions to the BA equations are classified by the
integers $Q_i, i= 1,\dots, M$ parameterizing the branches of the
logarithms. If one assumes that $Q_i=Q$ for all Bethe roots
then this quantum number gets shifted by one
at each RG cycle  and was identified with the integer
parameterizing the solution to the gap equations,

\begin{equation}
    \Delta_Q\propto \Delta_0 \exp^{-\lambda Q}.
    \label{delta_Q}
\end{equation}

At the
large $N$ limit the BA equations of the RD model reduce
to the BA equation of the Richardson-Gaudin model
with the rescaled coupling,
\beq
G_{Q}^{-1}= \eta^{-1}({\alpha} + \pi Q),
\eeq
which can be treated as the shifted boundary
condition in the generalized Gaudin model
parameterized by an integer.
Let us  emphasize that  the unusual cyclic RG behavior
is due to the presence of two couplings in the RD model.

\section{Triality in the integrable models and RG cycles}
\begin{figure}
    \centering
    \begin{equation*}
    \xymatrix{ \text{RS model}\ar@{<->}^{\text{QC duality}}[rr]\ar_{\begin{matrix}\text{non-relativistic}\\\text{limit}\end{matrix}}[dd]&&{\text{XXX chain}}\ar@{<->}[rr]\ar^{\begin{matrix}\text{semiclassical}\\\text{limit}\end{matrix}}[dd]&&\text{RD model}\\
    \\
    \text{Calogero system}\ar@{<->}^{\text{QC duality}}[rr]&&{\text{Gaudin model}}\ar@{<->}[rr]&&\text{Richardson model}
    }
    \label{triality}
    \end{equation*}
    \caption{Besides the triality shown on the picture, a bispectral duality acts on RS/Calogero and XXX/Gaudin sides of the correspondence. Being originated from three-dimensional mirror-symmetry \cite{gk}, this duality interchanges coordinates with Lax eigenvalues in the classical systems, and inhomogeneities with twists in quantum ones.}
    \label{fig:triality}
\end{figure}
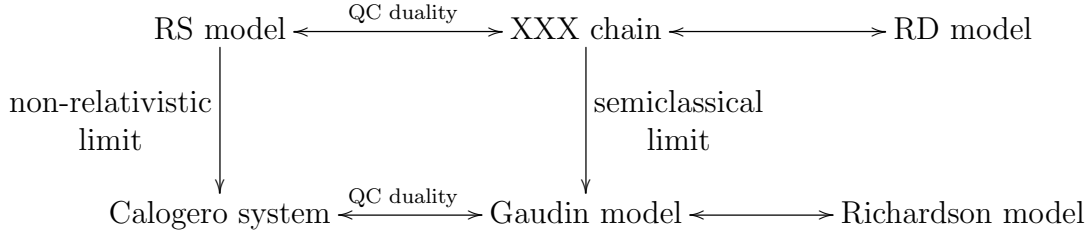

In this Section we summarize several dualities between the integrable models
and consider the realizations of the cyclic RG flows in these systems. The question
is motivated by the close relationship between the restricted BCS models and spin chains.
Actually there are three different families of models related with each other by the particular
identifications of phase spaces  and parameters. The first family  concerns the
system of fermions (Richardson-Russian Doll) which develop superconducting gap.
The second family involves the spin systems
of twisted inhomogeneous Gaudin-XXX-XXZ  type and their generalizations.
The third family
involves  the Calogero-Ruijsenaars (CR) chain of the integrable many body systems.

We look for the answers on the following questions
\begin{itemize}
\item What is the condition yielding the RG equation for some coupling in each family?
\item What is the RG variable?
\item What determines the period of the cycle?
\end{itemize}

In the superconducting system at RG step
one decouples the highest energy level and looks at the renormalization
of the interaction coupling constant. The RG time is identified with the
number of energy levels $t=\log N$. The period of  RG is
defined by the T-asymmetric parameter of RD model.

In the spin chain model the RG step corresponds to the "integrating out" one
"highest" inhomogeneity with the corresponding renormalization of the twist. The period of the
RG flow is fixed by the Planck constant in the quantum spin chain. In the bispectral
dual spin chain  \cite{bispectral} one now "integrates out" one of the twists and "renormalizes" the
inhomogeneity. Since the Planck constant gets inverted upon bispectrality
$\hbar_{spin} \rightarrow \hbar_{spin}^{-1}$ the period of the RG cycle
gets inverted as well. Note that the RG equation in the superconducting model
can be mapped into BAE in the spin chain \cite{lrs}. The condition yealding
the RG equation corresponds to the independence of the Bethe root on the
RG step.

 For two-body
system with attractive rational potential one can define the  RG condition
as the continuity of the zero-energy wave function under the changing the cutoff
scale at small $r$. This condition imposes the RG equation at the cutoff UV coupling
constant. This RG equation has the cycle with the period
\beq
T_{Cal}=\frac{\pi}{\nu-\frac12}.
\label{period}
\eeq
as was shown in Section \ref{RG_Cal}.

The Quantum-Classical (QC) duality \cite{gk, gzz} relates  the quantum spin chain systems
and the classical Calogero--type systems.
Through the QC correspondence, the rational Gaudin model can be linked with the rational Calogero system spin chain  inhomogeneities being the Calogero coordinates, and the twist in the spin chain  (which is a single variable in our case) being the Lax matrix eigenvalue.
It is also possible to make a bispectrality transformation of rational Calogero model, which interchanges Lax eigenvalues with coordinates. This means that now the Calogero coordinates correspond to the twists at the spin chain side. In this case the Calogero coupling gets inverted which means that the period of the RG cycle gets inverted as well.

To consider the mapping of RG cycles in the Calogero system and the spin chain we need the generalization of QC duality
to the quantum-quantum case. The spectral problem in Calogero model  has been identified with the KZ equation involving the Gaudin Hamiltonian,
\beq
\frac{d}{dz_i}\Psi =H_{gaud}\Psi + \lambda \Psi.
\eeq
Since we formulate RG condition on the Calogero side for the $E=0$ state, the inhomogeneous term in the
KZ equation is absent. The simplest test of the mapping of the RG cycles under QQ duality concerns the identifications
of the periods. On the spin chain side it is  identified with the Planck constant while at the Calogero side the period is defined by the coupling constant. The following identification holds for   QC duality  \cite{gzz}:
\beq
\hbar_{spin}= \nu,
\eeq
which implies that the periods of the cycles at the Calogero and spin chain sides match.

The Efimov-like tower in these families have the following interpretations.
In the superconducting system it corresponds to the family of the gaps $\Delta_n$ with
the Efimov scaling responsible of the scale symmetry broken
down to the discrete subgroup. In the spin chain it corresponds to the
different branches of the solutions to the BAE which can be also interpreted
in terms of the allowed set of twists.
Finally in the CR family it corresponds to the family of the shallow bound
states near the continuum threshold.


\section{ RG cycles in $\Omega$-deformed SUSY gauge theories}

In this Section we shall explain how the RG flows in $\Omega$- deformed SUSY
gauge theories can be reformulated
in terms of the brane moves. Why the very RG cycles could be expected in the
deformed gauge theories? The answer is based on the identification
of the quantum spin chains in one or another context in the SUSY gauge theory.
Once such quantum spin chain has been found we can apply the results
of the previous sections where the place of the RG cycles in the spin chain
framework has been clarified.

First, we shall briefly review the $\Omega$-deformation of the SUSY gauge theories.
Then we make some general comments concerning the realization of the gauge theories
as the worldvolume theories on $D$-branes to explain how the parameters of the
gauge theory are identified with the brane coordinates.

\subsection{Four-dimensional $\Omega$-deformed gauge theory}

The Bethe ansatz equations can be encountered not only in the models of superconductivity, but also in gauge theories. The quantum XXX spin chain governs the moduli space of vacua of an $\Omega$-deformed four-dimensional theory in the Nekrasov--Shatashvili limit, i.e. when one of the deformation is chosen to be zero: $\epsilon_2=0, \epsilon_1=\epsilon$ \cite{ns}. Since the quantum XXX spin chain displays a cyclic RG behaviour, as we have seen in the Section \ref{rg_chain}, it is interesting to identify this phenomenon in the four-dimensional gauge theory.

Consider a four-dimensional $\mathcal N=2$ theory with matter hypermultiplet, which has a vanishing $\beta$-function, i.e. when $N_f=2N_c$. This theory is dual to a classical inhomogeneous twisted XXX chain, in a sense that the Seiberg-Witten curve for the gauge theory coincides with the spectral curve for the spin chain. The twist of the spin chain is identified with the modular parameter of the curve and with the complexified coupling of the gauge theory, the inhomogeneities of the spin chain get mapped into masses of the hypermultiplets. For more information on the correspondence between classical integrable systems and gauge theories the reader can consult \cite{mironovgorsky}.

The $\Omega$-deformation is introduced to regularize the instanton divergence in the partition function of the gauge theory \cite{nekrasov}. We can consider the four-dimensional theory as a reduction of the six-dimensional $\mathcal N=1$ theory with metric:

\begin{equation}
  ds^2=2dzd\bar z+\left(dx^m+\Omega^{mn}x_nd\bar z+\bar \Omega^{mn}x_ndz  \right)^2,\qquad m=1,\ldots,4,
  \label{metric_omega}
\end{equation}
i.e. we can consider the theory on a four-dimensional space, fibered over a two-dimensional torus.
One can imagine the $\epsilon_{1,2}$ deformation parameters as chemical potentials for the rotations in two orthogonal planes in four-dimensional Euclidean space. One can also think that the Euclidean $\mathbb R^4$ space gets substituted by a sphere $S^4$ with finite volume.

The non-trivial $\Omega$-deformation modifies the correspondence between gauge theories and integrable systems. Namely, in the Nekrasov-Shatashvili limit the $\Omega$-deformed gauge theory corresponds to a quantum XXX spin chain with $\epsilon$ playing the role of the Planck constant \cite{ns}. This deformed gauge theory also appears to be dual to the two-dimensional effective theory on a worldsheet of a non-abelian string \cite{dorey}.

Consider $\Omega$-deformed $\mathcal N=2$ SQCD with $SU(L)$ gauge group, $L$ fundamental hypermultiplets with masses $m^f_i$ and $L$ antifundamental hypermultiplets with masses $m^{af}_i$.  Let us denote the set of the eigenvalues of the adjoint scalar in the vector multiplet by $\vec a$. We can expand the deformed partition function around $\epsilon=0$ to identify the prepotential and effective twisted superpotential,

\begin{equation}
    \log \mathcal Z\left( \vec a, \epsilon_1, \epsilon_2 \right)\sim \frac{1}{\epsilon_1 \epsilon_2}\mathcal F(\vec a, \epsilon)+\frac{1}{\epsilon_2}\mathcal W(\vec a,\epsilon).
    \label{log_Z}
\end{equation}

The effective twisted superpotential is a multivalued function, with the branch fixed by the set of integers $\vec k$:

\begin{equation}
    \mathcal W(\vec a, \epsilon)=\frac{1}{\epsilon} \mathcal F(\vec a, \epsilon)-2\pi i \vec k\cdot \vec a, \qquad \vec k\in \mathbb Z^L.
    \label{branch}
\end{equation}
The equation on vacua,

\begin{equation}
    \frac{\partial\mathcal W(\vec a, \epsilon)}{\partial \vec a}=\vec n, \qquad \vec n\in\mathbb Z^L,
    \label{f-term}
\end{equation}
provides the condition on $\vec a$,

\begin{equation}
    \vec a=\vec m^f-\epsilon \vec n.
    \label{a}
\end{equation}

This theory admits the existence of non-abelian strings probing the four-dimensional space-time. The two-dimensional worldsheet theory of the non-abelian string involves $L$ fundamental chiral multiplets with twisted masses $M^F_i$ and $L$ antifundamental multiplets with twisted masses $M^{AF}_i$, which are identified as:

\begin{equation}
    M^F_i=m^f_i-\frac 32 \epsilon, \qquad M^{AF}_i=m^{af}_i+\frac12 \epsilon.
    \label{masses}
\end{equation}
The two-dimensional theory also contains an adjoint chiral multiplet with mass $\epsilon$. The rank of the gauge group  $N$ (or equivalently the number of non-abelian strings) is given in terms of $\vec n$ vector by the relation:

\begin{equation}
    N+L=\sum_{l=1}^L n_l.
    \label{N_def}
\end{equation}

The modular parameters of the four-dimensional and the two-dimensional theories are related as:

\begin{equation}
    \tau_{2d}=\tau_{4d}+\frac{1}{2}(N+1).
    \label{tau}
\end{equation}

The effective twisted worldsheet superpotential is given in terms of the four-dimensional superpotential:

\begin{equation}
    \mathcal W_{4d}\left( a_i=m^f_i-\epsilon n_i,\epsilon \right)-\mathcal W_{4d}\left( a_i=m^f_i-\epsilon, \epsilon \right)=\mathcal W_{2d}\left(\{n_i\}\right).
    \label{w_2d}
\end{equation}

The two-dimensional superpotential depends on the set of eigenvalues of the adjoint scalar in vector representation $\lambda_i$, $i=1, \ldots, N$. The set of equations $\partial \mathcal W_{2d}/\partial \lambda=0$ appears to be equivalent to the Bethe ansatz equations for the XXX spin chain:

\begin{equation}
    \prod_{l=1}^L \left( \frac{\lambda_j-M^{F}_l}{\lambda_j-M^{AF}_l}\right)=\exp\left( 2\pi i \tau_{4d} \right) \prod_{k\neq j}^N\left( \frac{\lambda_j-\lambda_k-\epsilon}{\lambda_j-\lambda_k+\epsilon} \right).
    \label{bae_2d}
\end{equation}

The Planck constant in the spin chain is identified with the $\epsilon$ deformation parameter. The complexified coupling parameter plays the role of twist in the spin chain. The renormalization of the spin chain amounts to decoupling of one fundamental and one anti-fundamental chiral multiplet. In the four-dimensional theory it corresponds to the decrease of the number of flavors $N_f\to N_f-2$ simultaneously with reducing the rank of the gauge group $N_c\to N_c-1$. Therefore the theory remains conformal. The renormalization of the coupling constant analogous to (\ref{rg_discr}) derived from the relation (\ref{bae_2d}) for $N=1$ is:

\begin{equation}
  \exp\left(2\pi i( \tau_L-\tau_{L-1}) \right)=\frac{\lambda-M^F_L}{\lambda-M^{AF}_L}.
    \label{rg_2d}
\end{equation}

If we choose the masses to be equidistant with spacing $\delta m$, the change in the coupling constant during one step of RG flow is:

\begin{equation}
  \exp\left( 2\pi i(\tau_L-\tau_{L-1}) \right)\propto \frac{\epsilon}{\delta m}.
    \label{dm}
\end{equation}

Hence a number of nonperturbative scales emerges in a theory, analogously to the generation of the Efimov scaling in the Calogero model. These scales correspond to multiple gaps in the superconducting model:

\begin{equation}
    \Delta_n\propto \Delta_0 \exp\left(- \frac{\pi n \delta m}{\epsilon} \right).
    \label{delta_2d}
\end{equation}

Note that the emergence of cyclic RG evolution is a feature caused by the $\Omega$-deformation, since in a non-deformed theory a decoupling of the heavy flavor does not lead to any cyclic dynamics.

\subsection{$3d$ gauge theories and theories on the brane worldvolumes}

Let us briefly explain the main points concerning the
geometrical engineering of the gauge theories on
the $D$-branes suggesting the reader to consult the
details in the review paper \cite{giveon}. The $Dp$ brane
is the $(p+1)$-dimensional hypersurface  in the ten-dimensional space-time which
supports the $U(1)$ gauge field. This feature provides
the possibility to built up the gauge theories with the
desired properties. Let us summarize the key elements
of the "building procedure".

\begin{figure}
    \centering
    \includegraphics[height=70pt]{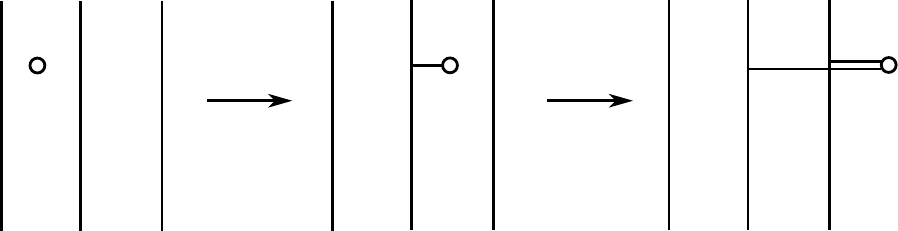}
    \caption{Hanany-Witten move. Here vertical lines are $NS5$ branes, horizontal lines are $D3$ branes, and circles are $D5$ branes. When a $D5$ brane is moved through a sequence of $NS5$ branes the linking number between them is conserved hence additional $D3$ branes appear.}
    \label{HWmove}
\end{figure}

\begin{itemize}

\item A stack of coinciding $N$ $D$-branes  supports $U(N)$ gauge theory with the maximal
supersymmetry.

\item Displacing some branes from the stack in the
transverse direction corresponds to the Higgs mechanism
in the $U(N)$ gauge theory and the distance between branes
corresponds to the Higgs vev.

\item To reduce the SUSY one imposes some boundary conditions
at some coordinates using other types of branes  or rotates
some branes.

\item All geometrical characteristics of the brane configurations
have the meaning of parameters of the gauge theory like
couplings or vevs of some operators in the
gauge theory on their worldvolumes.

\item If we move some brane through another one the brane
of smaller dimension could be created. The Hanany-Witten
move is the simplest example (see fig. \ref{HWmove}).

\item Since generically we have branes of different dimensions
in the configuration, for example, $N$ $D2$ branes and $M$ $D4$ branes
we have simultaneously $U(N)$ 2+1 dimensional gauge theory and
$U(M)$ dimensional 4+1 dimensional theory on the brane worldvolumes.
These theories coexist simultaneously hence there is highly nontrivial
interplay between two gauge theories.

\end{itemize}

Let us explain now how these brane rules can be used to engineer the gauge
theories which are related with the quantum spin chains. Our main example is a $3d$ $\mathcal N=2$ quiver gauge theory.

The brane configuration relevant for this theory is built as follows.
We have $M$ parallel  $NS5$ branes  extended in $(012456)$, $N_i$ $D3$ branes extended
in $(0123)$ between $i$-th and $(i+1)$-th $NS5$ branes, and $M_i$ $D5$ branes extended
in $(012789)$ between $i$-th and $(i+1)$-th $NS5$ branes (see table \ref{table:branes}). From this brane configuration
we obtain the $\prod_{i}^{M} U(N_i)$ gauge group on the $D3$ branes worldvolume with
$M_i$ fundamentals for the $i$-th gauge group. The distances between the $i$-th and $(i+1)$-th
$NS5$ branes yield the complexified gauge coupling for $U(N_i)$ gauge group while
the coordinates of the $D5$ branes in the $(45)$ plane correspond to the masses
of fundamentals. The positions of the $D3$ branes on $(45)$ plane correspond to the coordinates on the
Coulomb branch in the quiver theory.
The additional $\Omega$ deformation reduces the theory with $\mathcal N=4$ SUSY
to the $\mathcal N=2^*$ theory, i.e. an $\mathcal N=2$ theory with massive adjoint. It is identified as $3d$ gauge theory when the distance between $NS5$
is assumed to be small enough. We assume that one coordinate
is compact that is the theory lives on $\mathbb R^2\times S^1$.

\begin{figure}
\begin{tabular}{|c|c|c|c|c|c|c|c|c|c|c|}
    \hline
    &0&1&2&3&4&5&6&7&8&9\\
    \hline
    D3&$\times$&$\times$&$\times$&$\times$&&&&&&\\
    \hline
    NS5&$\times$&$\times$&$\times$&&$\times$&$\times$&$\times$&&&\\
    \hline
    D5&$\times$&$\times$&$\times$&&&&&$\times$&$\times$&$\times$\\
    \hline
\end{tabular}
\caption{Brane construction of the $3d$ quiver theory.}
\label{table:branes}
\end{figure}

The mapping of the gauge theory data into the integrability framework goes as follows. In the NS limit of the $\Omega$-deformation the twisted superpotential in $3d$ gauge theory
on the $D3$ branes gets mapped into the Yang-Yang function for the $XXZ$ chain \cite{ns}. The minimization of the superpotential yields the equations
describing the supersymmetric vacua and in the same time they are the Bethe ansatz equations for the $XXZ$ spin chain, generally speaking the nested Bethe ansatz equations. That is $D3$ branes
are identified with the Bethe roots which are distributed according to the ranks of the
gauge groups at each of $M$ steps of nesting $\prod_{i}^{M} U(N_i)$. The
positions of the $D5$ branes in the $(45)$ plane correspond to the inhomogeneities
in the $XXZ$ spin chain. The anisotropy of the $XXZ$ chain
is defined by the radius of the compact dimensions while the parameter of the $\Omega$
deformation plays the role of the Planck constant in the $XXZ$ spin chain. At small radius
the XXZ spin chain turns to the XXX spin chain. The twists in the spin chain correspond to the coordinates of the $NS5$ branes in the (78) plane, and the Fayet--Iliopoulos parameters in the three-dimensional theory \cite{gk}.

One step of the RG flow corresponds to elimination of one inhomogeneity in the spin chain resulting in renormalization of the twists. In the three-dimensional theory this means that the integration of one massive flavor leads to the renormalization of the FI parameters. In terms of the transformations of the brane configurations this process receives transparent geometrical interpretation:

\begin{itemize}

\item The RG step is the removing of one $D5$ brane which amounts to the renormalization of the position of $NS5$ branes or twists.

\item The period of the RG cycle is fixed by the number of $NS5$ branes \cite{bg}, since
it was identified with the Planck constant in the spin chain.

\item At some scale  the twists flow from $+\infty$
to $-\infty$.

\end{itemize}

\section{Conclusion}
Are there any general lessons which we could learn for
the quantum field theory from the very existence  of the
cyclic RG flows? The most
important point is that there is some
fine structure at the UV scale which is reflected in the Efimov
tower with the BKT scaling behavior. Moreover  the cyclic
flows imply the interplay between the UV and IR cutoffs in the theory
which usually was attributed to the noncommutative theory. This
mixing presumably could shed the additional light on the dimensional
transmutation phenomena in the field theory and provide the examples
for the simultaneous generation of the multiple scales.

The presence
of two parameters in RG is quite common however probably some
additional properties of these parameters are required. In particular
in many (although not all) examples the period of the cycle is fixed by the
``filling fraction'' in some external field which could be magnetic
field or parameter of $\Omega$ background. The latter has the meaning
of the Planck constant in the auxiliary finite dimensional integrable model.
This could suggest that the very issue can be formulated purely
in terms  of the quantum phase space since the Planck constant
can be interpreted as the external field applied to the classical
phase space.

Actually we could expect the relation of RG cycles with some
refinement of the path integral in quantum mechanics. As an aside
remark note that the attempt to get the rigorous mathematical
formulation of the renormalization of the QFT leaded 
to the motivic generalization of the path integral. It corresponds
to some fine structure at the regulator scale which has some
similarities with the discussion above. The RG cycle in the
quantum rational Calogero model implies the intimate relation with the knot
theory since the knot invariants at the rational Calogero coupling
are the characteristics of the Calogero spectrum (cf. \cite{bg}).

As we already mentioned, cyclic renormalization dynamics is connected with BKT--pairing of partons in two-dimensional model. One could wonder whether this connection is universal. One four-dimensional example of such pairing has to be mentioned. It is bion condensation in 3+1 dimensions. The RG analysis of the model involving the gas of bions and
electrically charged  W-bosons has been considered in \cite{unsal2}
where the RG flows involves the fugacities for electric and magnetic
components and the coupling constant. The coupled set of the
RG equations has been solved explicitly in the self-dual case and the solution
to the RG equations for the fugacities obtained in  \cite{unsal2}
is identical to the solution for the coupling in the RD model
upon the analytic continuation. The period of the RG in the solution above is fixed
by the RG invariant which has been identified with the product
of the UV values of the electric and magnetic fugacities $y_e\times y_m$.
The similarity between the RG behavior is not accidental
since the mapping of the gauge theory and the perturbed
XY model has been found in \cite{unsal2}.

We would like to emphasize that the investigation of various aspects of limit cycles in RG dynamics still remains on its early stage and there is a considerable number of open questions. The RG cycles can have numerous applications to different aspects of mathematical physics. In this case the RG dynamics is considered as an example of non-trivial dynamical system.

The work of A.G. and K.B. was supported in part
by grants RFBR-12-02-00284 and PICS-12-02-91052. The work of K.B. was also supported by the ``Dynasty'' fellowship. A.G. thanks
FTPI at University of Minnesota where the part of this work has been done
for the hospitality and support.
We would like to thank N. Nekrasov and F. Popov for useful discussions and comments.

\end{document}